# Studying the Similarity of COVID-19 Sounds based on Correlation Analysis of MFCC

Mohamed Bader[1], Ismail Shahin[2], Abdelfatah Hassan[3]
Department of Electrical Engineering
University of Sharjah
Sharjah, United Arab Emirates
[1]U16104773@sharjah.ac.ae, [2]ismail@sharjah.ac.ae, [3]U17200027@sharjah.ac.ae

*Abstract*— Recently there has been a formidable work which has been put up from the people who are working in the frontlines such as hospitals, clinics, and labs alongside researchers and scientists who are also putting tremendous efforts in the fight against COVID-19 pandemic. Due to the preposterous spread of the virus, the integration of the artificial intelligence has taken a considerable part in the health sector, by implementing the fundamentals of Automatic Speech Recognition (ASR) and deep learning algorithms. In this paper, we illustrate the importance of speech signal processing in the extraction of the Mel-Frequency Cepstral Coefficients (MFCCs) of the COVID-19 and non-COVID-19 samples and find their relationship using Pearson's correlation coefficients. Our results show high similarity in MFCCs between different COVID-19 cough and breathing sounds, while MFCC of voice is more robust between COVID-19 and non-COVID-19 samples. Moreover, our results are preliminary, and there is a possibility to exclude the voices of COVID-19 patients from further processing in diagnosing the disease.

## I. INTRODUCTION

Since the horrific outbreak of the COVID-19 has occurred and was declared a global pandemic from the world health organization in March 2020, it has threatened almost all the human beings' lives on earth. For the time being, there are more than 11 million confirmed cases of COVID-19 infections in more than 200 countries. Based on the world health organization (WHO) statistics, there are 528,101 people who passed away after infected by the virus [1]. The new coronavirus is attracting a widespread interest due to its fast dissemination among the people and a vital impact on people with a weak immune system. Besides, the world health organization (WHO) has specified the vital symptoms of COVID-19 as high body temperature, muscle aches, and difficulties regarding breathing and coughing. By the escalation, low abilities to control the spread of the virus and not having the privilege of screening and testing [2]. Due to that, an innovative, effective, and modern way solution is ready to be implemented and integrated into the health sector. Lately, Artificial Intelligence (AI) has been widely implemented in the digital health sector. AI has many applications in the field of speech and audio analysis; it could be implemented in the screening and early detection of the infected people process, which could help control and reduce the number of infected people. It is stated that there is a relationship between the COVID-19 symptoms and a person's speech signal. Thus, it is possible to determine whether the person has an infection or not by performing speech analysis; in addition, the implementation of AI in design chatbots and programmed software would support whoever is suffering from the fallouts of the lockdown and quarantine.

This paper is organized as follows: Section II covers the literature review of ASR uses in digital health. Section III explains our methodology. Section IV discusses the experimental results. Finally, section V gives the concluding remarks of this work.

## II. LITERATURE REVIEW

AI can be implemented in the digital health sector for the diagnoses and early detection of the symptoms of COVID-19 based on the analysis of the cough, breath, and voice. In addition, it could be utilized to track the mental state of the patients who might be suffering from the aftermath of lockdowns and quarantine [3]. The detection of patients' symptoms severity, physical and mental state can be obtained by analyzing their speech recordings. Moreover, a low cost and a reliable health state detection software can be established by monitoring and analyzing the sleep-quality, severity of illness, fatigue, and anxiety [4]. The health condition of human beings and their mental state can be obtained from the analysis of the features of sounds. Diseases that are related to the respiratory system can be detected using the machine learning algorithms analysis. Moreover, differentiating among coughs to determine the type of disease can be done by evaluating the acoustic features using several classifiers such as Convolutional Neural Network (CNN) and Long Short Term Memory (LSTM) [5]. Besides, Voice Activity Detection (VAD) has been done based on the Mel-Frequency Cepstral Coefficients (MFCC) similarity, by collecting the correlation coefficients. It has shown that finding the similarity of the audio sound in a noisy background using MFCC is the most suitable method compared to other features which were evaluated [6]. An algorithm for matching patterns, recognizing and differentiating speech is developed based on using the MFCC as the extracted features and the principle of collecting the correlation coefficients [7]. Also, continues or heavy cough can be considered as a sign or indication for some kind of respiratory disease, thus tracking the condition and health state of patients can be done by implementing the auto cough detection by evaluating the acoustic features using deep learning algorithms. Features extraction such as Short-Time Fourier Transform (STFT), Mel Filter-Bank (MFB), and MFCC have been evaluated using classifiers such as CNN and LSTM. Moreover, the differentiation among coughs has been made based upon the acoustic features [8].

In this paper, we propose, analyze and study the evaluation of the MFCC acoustic features and the correlation analysis of these features, which were extracted from infected patients and healthy individuals to illustrate if there is a relationship or not. Also, an initial hypothesis would be provided on what symptoms should be considered best for tracking, monitoring, and diagnosing.

### III. METHODOLOGY

#### A. Data Collection

In this work, collecting data is considered as the first step. Presently, data gathering is underway globally from both contaminated patients and healthy individuals. Thus, our dataset is comprised of 7 healthy speakers (4 male and 3 female), and 7 COVID-19 infected patients (5 male and 2 female). Data of COVID-19 infected patients was collected from a hospital in Sharjah, United Arab Emirates. Each speaker was asked to cough four times, to take a deep breath, and to count from one to ten. Furthermore, the patients must sit with their head upright in a relaxed manner while recording their speech signals. As a result, three recordings per speaker were acquired within the data collection session using a mobile, which can affect the quality of the sound. Besides, data of healthy speakers were also collected from the United Arab Emirates speakers using the same mechanism. The total collected number of samples used in this study was 42 ((7 healthy speakers × 3 recordings) + (7 infected speakers × 3 recordings)). Due to the inconvenience caused by the epidemic, we were only able to get a proper dataset categorized either as healthy or COVID-19 samples.

#### B. Speech Pre-processing

Speech signals pre-processing is seen as a significant step after capturing the database. To increase the performance of our analysis, we need to perform pre-processing for our recordings. This process is done by isolating the part of the captured sound from silence [9]. Furthermore, assume that x(n) is a speech signal before the pre-processing, and s(n) is the clean speech signal, with a noise denoted by d(n). Then our speech signal before the preprocessing can be represented by the following expression,

$$x(n) = s(n) + d(n) \quad (1)$$

PRAAT is the program we used to complete Pre-Processing. PRAAT is a software tool that is designed to interpret and modify voice signals. It has several characteristics; however, we aim to eliminate portions of the silence. Therefore, we have to cut the silence portion at the beginning and at the end of the captured recordings.

#### C. Extraction of Features

Our speech signals have a set of information. The determination of this information is an essential task. Thus, the efficiency of this phase is vital for our analysis. In this work, the features extracted for the best parametric representation of acoustic signals are called Mel-Frequency Cepstral Coefficients (MFCCs) [9]. We have various features that can be extracted from the collected recordings. However, we only focused on extracting MFCCs since they are considered the most important features in distinguishing COVID from non-COVID sounds [5]. Also, MFCC is commonly used in speaker and emotion recognition [11], [12] [13], [14], [15]. Furthermore, MFCC relies on human listening features that cannot realize the frequencies above 1KHz. Therefore, signals are expressed in MEL scale, which utilizes one low-frequency filter below 1000 Hz, and another high-frequency filter above 1000 Hz [16], [17]. The computation steps of the MFCC is clarified in Fig. 1.

Mel frequency has been computed using the following six steps [6], [18], [19]:

1. **Pre-emphasis:** In this step the speech signal is passed to a high-pass filter. We can represent the phase pre-emphasis by,

$$y(n) = x(n) - a * x(n - 1) \quad (2)$$

where x(n) is the input speech signal, the output signal is referred to as y(n), and the value of $a$ is approximately between 0.9 and 1.0. The purpose of this step is to increase the energy in the signal at higher frequencies.

2. **Framing:** Is the process of dividing the speech signal that has $N$ samples into segments within the range of 20 ms to 40 ms. Adjoining frames are separated by $M$ (M<N). Typical values used are M = 100 and N= 256.

3. **Windowing:** Each frame will be multiplied by a Hamming window. This step will allow us to increase the continuity of the frames. We can represent the windowing with the following expression,

$$y(n) = x(n) * w(n) \quad (3)$$

where,

$$w(n) = 0.54 - 0.46 \cos\left(\frac{2\pi n}{N-1}\right) \quad (4)$$

4. **Fast Fourier Transform:** In this step, every frame that we obtained from the Hamming windowing is represented in the frequency domain energy distribution. We have to apply the Fourier Transform to get a better observation of speech signal characteristics.

5. **Mel Filter Bank:** The spectrum that we obtained has a wide range of frequencies. Therefore, we need to pass the spectrum to Mel Filter Bank, which consists of a set of triangular bandpass filters. In our analysis, the Mel Filter Bank contains a group of 25 triangular filters. Moreover, to compute the Mel frequency $m$, we can use the following expression,

$$m = 2595 \, log(1 + \frac{f}{100}) \quad (5)$$

6. **Discrete Cosine Transform (DCT):** The last step is to represent the Mel spectrum in the time domain to acquire the MFCCs. The set of coefficients is called acoustic vectors. Thus, every input speech signal is converted into an acoustic vector sequence. We can express DCT by the following formula,

$$C(n) = \sum_{K=1}^{N} cos[n * (k - 0.5) * \left(\frac{\pi}{N}\right)] E_K \quad (6)$$

In this work, the result of extracting MFCC features is an M x N matrix. Each $N$ column represents a signal frame, and each $M$ row represents the MFCC of the specific frame. From the collected recordings, we obtained 13 MFCCs. However, since the DCT step keeps the important speech information in the first few coefficients, we only utilized the first three obtained MFCCs and neglected the rest.

D. *Correlation Coefficients*

Correlation is the process of finding the degree of similarity between two variables. The correlation coefficient is a numerical measurement to quantify such a degree of similarity between variables. Pearson's correlation coefficients are the most commonly used in defining the degree of relationship between two objects [20]. We can calculate such a coefficient (R) by the following formula,

$$R = \frac{n(\sum XY) - (\sum X).(\sum Y)}{\sqrt{n(\sum X^2) - (\sum X)^2} \sqrt{n(\sum Y^2) - (\sum Y)^2}} \quad (7)$$

where *X* and *Y* are the measures of variables, and the Number of observations is referred to *n*. Moreover, correlation coefficients range between -1 and 1. Thus, to interpret the correlation coefficients, we can consider that two variables have an intense linear closeness when R is between 0.5 and 1. However, the range within 0 and 0.5, shows how the relationship between two variables is weak [6].

IV. RESULTS AND DISCUSSION

A. *Correlation analysis in time and frequency domain*

Two sound samples representing a voice for both COVID-19 infected patients and non-COVID-19 people have been tested in both the time and frequency domain to obtain the correlation coefficient, as shown in Fig. 2a and Fig. 2b respectively. Furthermore, the test has been made using MATLAB, showing unreasonable and unclear results. Thus, computing the correlation coefficients is not as effective as it could be while using this method.

B. *Correlation analysis of MFCC*

In this section, correlation tests have been made to obtain the correlation coefficients based on, namely, cough sounds, breath sounds, and voices, between two COVID-19 samples and between COVID-19 and non-COVID-19 samples. Two tests were made using python to obtain the correlation coefficients and represent them through the correlation matrix. Table I shows the average and the variance of the correlation coefficients of both tests. The first test was made between COVID-19 patients and healthy people samples of cough sound, breathing sound, and voices. Whilst, the second test was done among two COVID-19 samples. Fig. 3 and Fig. 4 illustrates the correlation matrix of both tests. The dark color on the correlation matrix indicates a strong correlation, and the light color indicates a weak correlation. Furthermore, it can be seen that the MFCCs of voice samples of both tests have a strong correlation. Whereas, cough and breathing sounds have an intense linear closeness for the second test only. It seems from our analysis that we can rely on extracting the MFCC of cough and breathing sounds to do a diagnostic system and we do not lean on patients' speech signals.

V. CONCLUDING REMARKS

This paper has expressed an innovative and modern solution for screening an early detection of the COVID-19 symptoms. It also illustrates the mechanism of correlation analysis in showing the similarity of speech features between COVID-19 samples. This is done by evaluating the MFCC acoustic features and obtaining the correlation coefficients. Based on the results from Table I, it can be stated that the voice of patients has shown a high correlation between healthy samples and infected samples which is not acceptable and in-efficient. The reason behind these results could be due to the small COVID-19 samples that have been collected, or because collecting the data through a mobile affects our results. Furthermore, these results can be useful in terms of using MFCC. For example, designing a diagnostic tool to determine whether the patient is infected or not. Moreover, these results are preliminary, and we can improve the functionality of our analysis by increasing the number of samples. we can conclude that tracking and analyzing the coughing and breathing of the patient is the most suitable way to detect infection. Due to the aftermath effects that some people might experience from the lockdown and quarantine.

ACKNOWLEDGMENTS

The authors would like to thank the University of Sharjah in the United Arab Emirates to fund this work through the competitive research project entitled "Emirati-Accented Speaker and Emotion Recognition Based on Deep Neural Network, No. 19020403139" and the spotlight project called "Capturing Emirati-Accented Speech Corpora for Applications of Speech Signal Processing" which is supported by the College of Engineering.

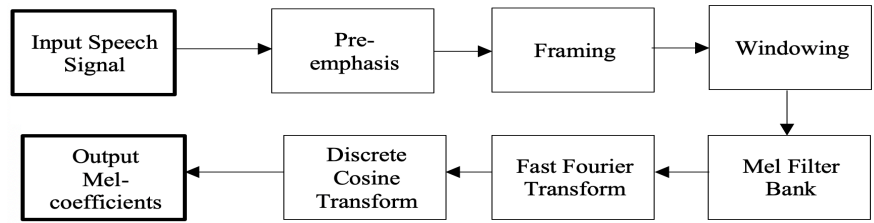

**Figure 1.** Block diagram of the MFCC algorithm

TABLE I. AVERAGE AND VARIANCE OF THE CORRELATION COEFFICIENTS OF BOTH TESTS.

| Test | Samples | Average correlation coefficients | Variance of the correlation coefficients | Correlation strength |
|---|---|---|---|---|
| Non-COVID-19 vs COVID-19 | Cough | 0.42 | 0.08 | Low positive correlation |
| | Breath | 0.43 | 0.08 | Low positive correlation |
| | Voice | 0.79 | 0.02 | High positive correlation |
| Covid-19 vs Covid-19 | Cough | 0.65 | 0.07 | Moderate positive correlation |
| | Breath | 0.58 | 0.1 | Moderate positive correlation |
| | Voice | 0.82 | 0.02 | High positive correlation |

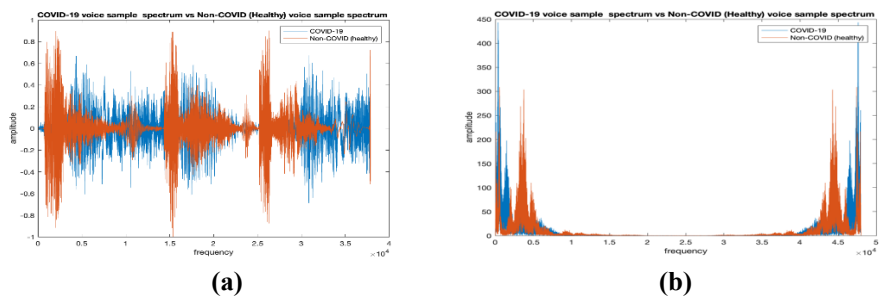

(a)      (b)

**Figure 2.** COVID-19 and non-COVID-19 voice signal (a) waveform in the time domain; (b) spectrum

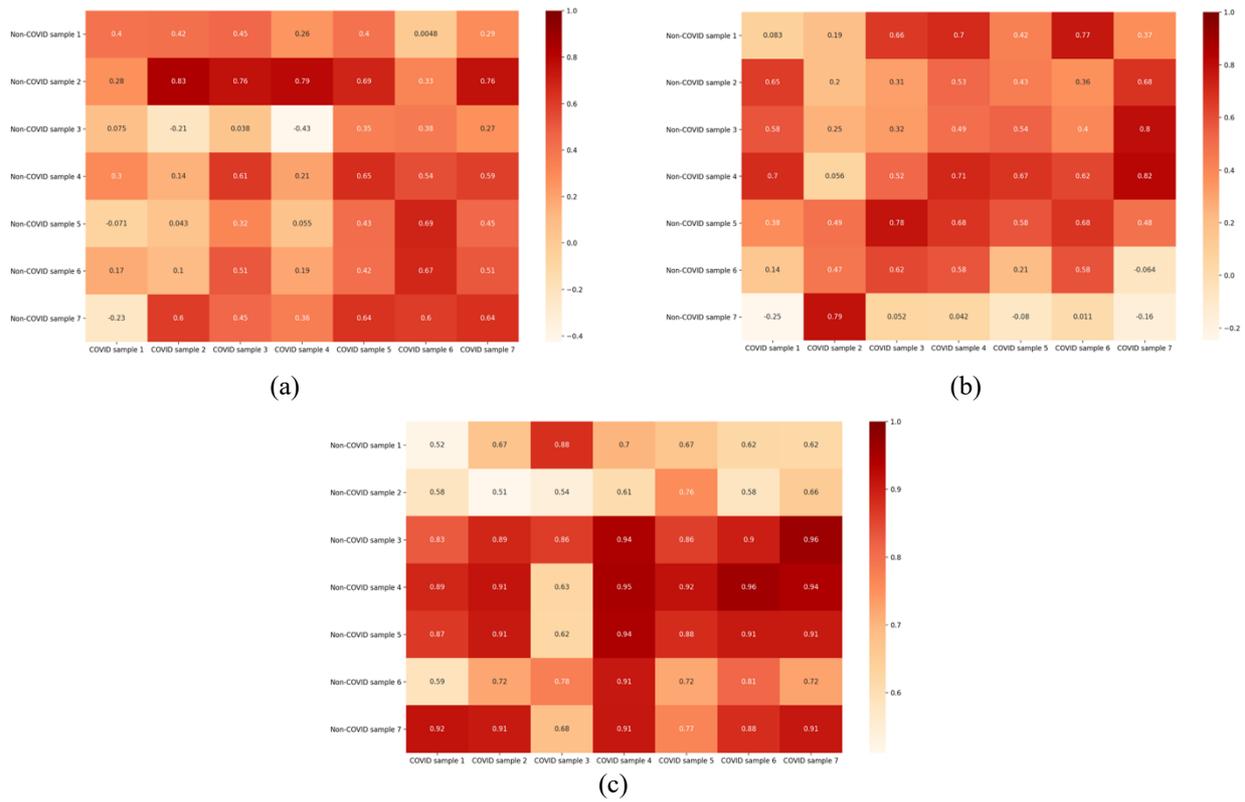

**Figure 3.** The correlation matrix between non-COVID-19 and COVID-19 samples of (a) cough sound; (b) breathing sound; (c) voice

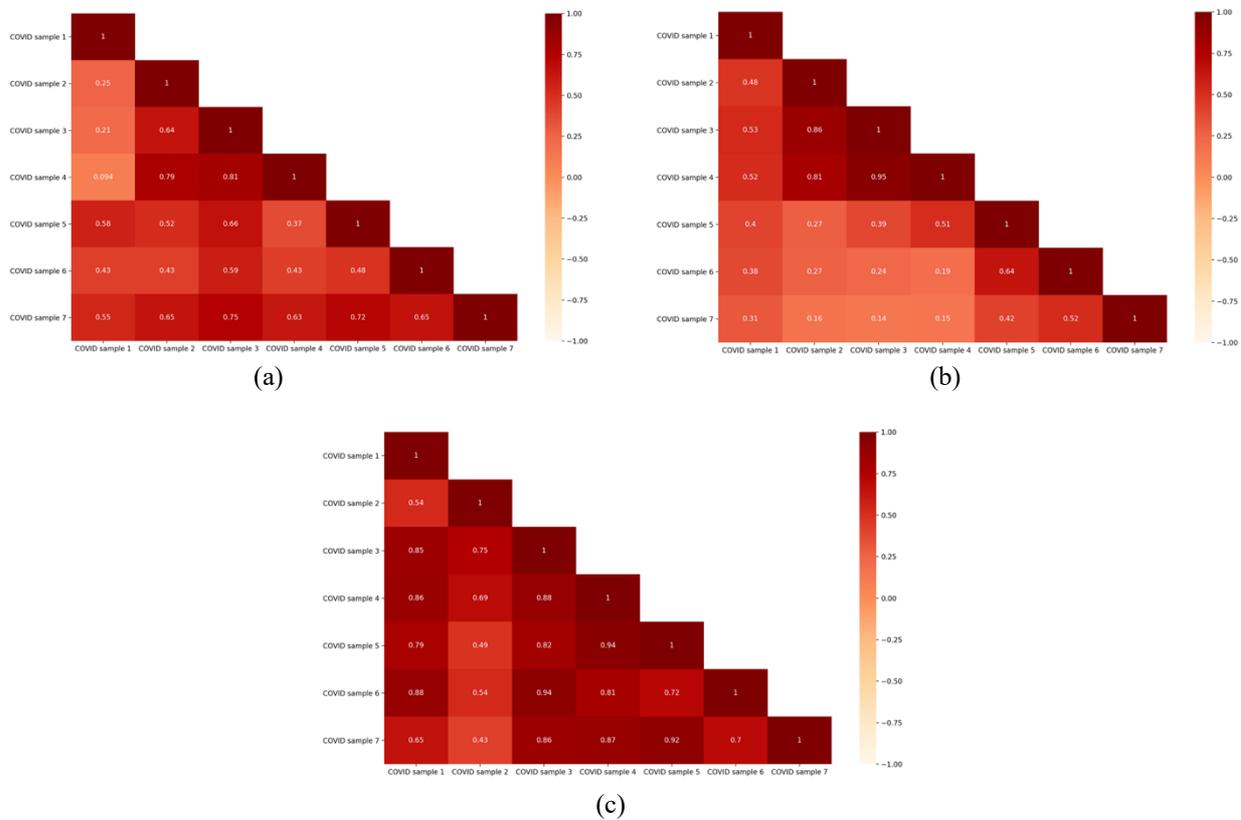

**Figure 4.** The correlation matrix between two COVID-19 samples of (a) cough sound; (b) breathing sound; (c) voice